\documentstyle[psfig]{mn}

\def\ltsima{$\; \buildrel < \over \sim \;$}
\def\lsim{\lower.5ex\hbox{\ltsima}}
\def\gtsima{$\; \buildrel > \over \sim \;$}
\def\gsim{\lower.5ex\hbox{\gtsima}}

\begin{document}

\title[Compton dragged gamma--ray bursts: the spectrum]
{Compton dragged gamma--ray bursts: the spectrum}
\author[Ghisellini, Lazzati, Celotti and Rees]
{Gabriele Ghisellini$^1$, Davide Lazzati,$^{1,2}$
Annalisa Celotti$^3$ and Martin J. Rees$^{4}$\\ 
$^1$ Osservatorio Astronomico di Brera, Via Bianchi 46, I--23807
Merate (Lc), Italy \\
$^2$ Dipartimento di Fisica, Universit\`a degli Studi di Milano,
Via Celoria 16, I--20133 Milano, Italy \\
$^3$ SISSA, Via Beirut 2--4, I--34014 Trieste, Italy \\
$^4$ Institute of Astronomy, Madingley Road, Cambridge\\
E--mail: {\tt gabriele@merate.mi.astro.it}, {\tt lazzati@merate.mi.astro.it}, 
{\tt celotti@sissa.it}, {\tt mjr@ast.cam.ac.uk}
}

\maketitle

\begin{abstract}
We calculate the spectrum resulting from the interaction of a 
fireball with ambient soft photons.
These photons are assumed to be produced by the walls of a funnel in a
massive star.
By parameterizing the radial dependence of the funnel temperature we
calculate the deceleration of the fireball self--consistently, taking
into account the absorption of high energy $\gamma$--rays due to
interaction with the softer ambient photons.
The resulting spectrum is peaked at energies in agreement with
observations, has a $\nu^2$ slope in the X--ray band and a steep
power--law high energy tail.
\end{abstract}


\section{Introduction}

We have recently proposed (Lazzati et al. 2000, hereafter Paper I)
that the gamma--ray burst (GRB) phenomenon originates from the
interaction of a relativistic fireball with a dense photon
environment, leading to Compton drag.
On one hand this is an inevitable effect if the progenitors of GRBs
are massive stars which are about to explode or have just exploded as
supernovae; on the other hand this mechanism greatly alleviates the
efficiency problem faced by the standard internal shock scenario
(Lazzati, Ghisellini \& Celotti 1999; Panaitescu, Spada \& Meszaros
1999; Kumar 1999).
In Paper I we have discussed the basic Compton drag scenario, showing
how this process can convert bulk motion energy directly into
radiation with a remarkable high efficiency and, on the basis of
simple estimates, how the resulting spectrum should peak, in a $\nu
F(\nu)$ representation, around $\sim$1 MeV, as observed.

Here we quantitatively and self--consistently estimate the predicted
spectrum, assuming that the fireball propagates in a funnel inside a
massive star, and show that, independently of the details of the model, 
it satisfactorily resembles what observed. 
Since the funnel walls emit a blackbody spectrum and the scattered
photons are boosted by the square of the Lorentz factor ($\Gamma$) of
the fireball, the local spectrum has a blackbody shape, at a
temperature enhanced by $\Gamma^2$.
However, the observed spectrum, convolution of all the locally emitted
spectra, is not a blackbody, due to four main effects:
i) the funnel walls would not be at a uniform temperature, but there should
be a gradient between the internal and external parts;
ii) if the Compton drag process is efficient, the fireball decelerates;
iii) the very high energy emission produced in the internal regions
can interact with the ambient photons, producing electron--positron pairs;
iv) the fireball may become optically thin to scattering outside the funnel,
where the ambient photons are characterized by the same 
temperature, but their energy density is progressively diluted with distance. 

\section{Basic assumptions}

We postulate that the fireball propagates with a bulk Lorentz factor
$\Gamma$ inside a funnel cavity, whose walls emit blackbody radiation
at a temperature $T$, of conical shape with semi--aperture angle
$\psi$.
The calculation starts at the distance $z_0$, assumed to be the end of
the acceleration phase and, for consistency, we verify that the power
emitted at $z<z_0$ is negligible.

We assume that the fireball is and remains cold
in the comoving frame.
At $z_0$, in fact, the internal energy has been already used
to accelerate the fireball, and thus protons are
sub--relativistic. On the other hand leptons might be still hot 
at $z_0$ and/or being re-heated 
when the bulk scattering process starts to be efficient. However, 
in a few (Compton) cooling timescales they would 
reach the (sub-relativistic) Compton temperature. It is thus 
reasonable to treat also the leptonic component as cold
in the estimate of both the dynamics and resulting spectrum.

The initial Lorentz factor (at $z_0$) is indicated as $\Gamma_0$, and the 
fireball energy is therefore $E_f=\Gamma_0 M_f c^2$, 
where $M_f$ is its rest mass.
For simplicity, the dependence of the temperature on $z$,
between $z_0$ and the radius of the star $z_*$,
has been parameterized by a power law:
\begin{equation}
T(z)\, =\, T_0 \left( z\over z_0 \right)^{-b}\, =\,
T_* \left( z\over z_* \right)^{-b}
\end{equation}
where $T_*$ is the temperature at the top of the funnel.

Inside it, we approximate the local radiation energy density 
of the ambient photons as $U(z)= a T^4(z)$.
Beyond $z_*$, and in the region where the fireball remains optically
thick (i.e. for $z<z_T$, see below), $U(z)$ is characterized by
the same temperature, but decreases. 
As the relevant quantity is 
the amount of radiation which is indeed scattered by the fireball, 
we parameterize the dependence on $z$ of the 
product $U(z) \times$ (the scattering rate) as $(z/z_*)^{-g}$.

We consider $g$ a free parameter. 
Inside the funnel $g=0$, while outside it 
a value $g>2$ can account for a decrease in the scattering rate
due to the changing of the typical scattering angle (photons come
preferentially at smaller angles as $z$ increases).
As the scattering rate is $\propto (1-\beta\cos\theta)$, where $\theta$ 
is the angle between the photon and the electron directions,
far from the star surface $(1-\beta\cos\theta) \propto
(z/z_*)^{-2}$, corresponding to $g\sim 4$.
Furthermore some of the radiation produced by the massive star 
could be reflected and re--isotropized by scattering material, of unknown 
radial density profile, likely surrounding the massive star progenitor. 
In particular if this forms a wind with a $z^{-2}$ profile, 
the energy density of the re--isotropized radiation scales as $z^{-3}$,
and dominates the seed photon distribution at large distances.
In this case 
$U(z)\times$(the scattering rate)
 can have a complex profile,
being flat in the vicinity of the surface of the star, then decreasing
as $z^{-2}$ and as $z^{-4}$ for increasing $z$,
to become flatter when the component associated with the
re--isotropized photons dominates.
It is also possible that, as a result of intermittent stellar activity,
the stellar wind is not continuous.
In this case a single shell may dominate the scattering, producing a
homogeneous and isotropic scattered radiation field, dominating
the total radiation energy density beyond some critical distance.

The distance $z_T$ at which the fireball becomes optically thin to
scattering is
\begin{equation}
z_T = \left({\sigma_T E_f \over \pi \psi^2 m_p c^2 \Gamma_0}\right)^{1/2}
\sim 
3.7\times 10^{14} \psi_{-1}^{-1}
E_{f,51}^{1/2} \Gamma_{0,2}^{-1/2}  {\rm cm}
\end{equation}
where the conventional representation $Q = Q_x 10^x$ and c.g.s. units 
are adopted. 
It is then likely that the fireball becomes transparent at $z>z_*$ 
(since the radius of red supergiants is $z_*\lsim 10^{13}$ cm).

As long as the fireball is opaque to scattering, the interaction 
with photons boosts their energy by a factor $\sim 2\Gamma^2$.
Therefore the (local) total energy emitted by
the fireball through the Compton drag process (over a distance $dz$) is
\begin{equation}
dE(z) \, = \, 2 \pi \psi^2 z^2 aT_0^4\left({z\over z_0}\right)^{-4b} 
\Gamma^2 dz \quad z<z_*
\end{equation}
\begin{equation}
dE(z) \, = \, 2 \pi \psi^2 z^2 aT_*^4 
\left({ z\over z_*}\right)^{-g}\, \Gamma^2 dz \quad z>z_*.
\end{equation}
The factor 2 in front of the RHS of these equations takes into account
that the preferred scattering angle is $\sim 90^\circ$, corresponding
to an average energy boost of $2\Gamma^2$.

Let us now consider the spectral shape.
For this it is convenient to use dimensionless photon energies and
temperatures, defined as $x\equiv h\nu/(m_e c^2)$ and $\Theta \equiv
kT/(m_ec^2)$, respectively.

The resulting Compton spectrum has a blackbody shape, of effective
temperature $T_c = 2\Gamma^2 T$ (or $\Theta_c = 2\Gamma^2 \Theta$),
i.e.  the local spectral distribution produced within $dz$ is given by:
\begin{eqnarray}
dE(z, x)&=&\pi^2 \psi^2 {z^2 \over \Gamma^6}
m_ec^2 \left({m_e c\over h}\right)^3  { x^3 \over e^{x/\Theta_c}-1} dz; \,
\nonumber \\ &\, &  \qquad \qquad  \qquad\qquad \qquad\qquad z<z_*
\end{eqnarray}
\begin{eqnarray}
dE(z, x)&=&\pi^2 \psi^2 {z^2 (z/z_*)^{-g}\over \Gamma^6} 
m_ec^2 \left({m_e c\over h}\right)^3
{ x^3 \over e^{x/\Theta_{c,*}}-1} dz; \nonumber \\ 
&\, &  \qquad \qquad  \qquad\qquad \qquad\qquad  z>z_*,
\end{eqnarray}
where $\Theta_{c,*}=2\Gamma^2\Theta_*$.
Equations (5) and (6) are correctly normalized, i.e. the integrated 
energies correspond to those expressed in (3) and (4).

\section{The fireball dynamics}

As long as the fireball remains optically thick for scattering
and this occurs in the Thomson regime, the dynamics (deceleration) 
of the fireball due to the radiative drag, obeys: 
\begin{equation}
M_f c^2 \, {d\Gamma \over dz}\, = \, -\,  2\pi \psi^2 z^2  aT^4 \Gamma^2.
\end{equation}
Assuming the temperature profile of equation (1) we obtain:
\begin{eqnarray}
\Gamma &=& {\Gamma_0 \over 1+ 2\pi \psi^2 aT_0^4 \Gamma_0^2 z_0^3 
[(z/z_0)^{3-4b}-1]/[E_f(3-4b)]};  
\nonumber \\ &\, &  \qquad \qquad  \qquad\qquad \qquad\qquad z_0<z<z_*, 
\end{eqnarray}
and thus the deceleration radius, $z_d$, defined as the distance at which 
$\Gamma$ is halved, corresponds to:
\begin{eqnarray}
z_d\, &=&\, 
z_o\, \left[ 1+ { E_f (3-4b) \over 
2\pi\psi^2 aT_0^4\Gamma_0^2 z_0^3}\right]^{1/(3-4b)}\, \nonumber \\
&=& \, 
z_o\, \left[ 1+ { E_f (3-4b) \over 
2\pi\psi^2 aT_*^4\Gamma_0^2 (z_*/z_0)^{4b}z_0^3}\right]^{1/(3-4b)}
\end{eqnarray}
Beyond $z_d$, the Lorentz factor decreases with distance as a power
law, whose slope is determined by the temperature profile.

Outside the star radius ($z>z_*$) the Lorentz factor follows:
\begin{eqnarray}
\Gamma &=& {\Gamma_* \over 1+ 2\pi \psi^2 aT_*^4 \Gamma_0\Gamma_* z_*^3 
[(z/z_*)^{3-g}-1]/[E_f(3-g)]};
\nonumber \\ &\, &  \qquad \qquad  \qquad\qquad \qquad\qquad \,\, z_*<z<z_T.
\end{eqnarray}
Note that Klein-Nishina effects are important for incoming photon energies  
such that $x \Gamma>1$, i.e. when $\Theta>1/(3\Gamma)$.
For simplicity, we neglect interactions in this regime when
calculating $\Gamma(z)$, but we assume no scattering events when
$\Theta>1/(3\Gamma)$ in calculating the spectrum.
This simplification is justified as long as most of the fireball energy
is lost in the Thomson scattering regime (see Fig. 2, which shows
that $\Gamma$ starts to decrease at distances where the temperature
is small enough to ensure scatterings entirely in the Thomson regime).

When the fireball becomes optically thin, the amount of scattered
photons is correspondingly reduced, and the process becomes
less efficient. 
As shown by equation (2), this is likely to happen at some
distance from the star surface, where the photon density is also
reduced, thus further decreasing the efficiency of the process.
In the numerical calculations we have however included the optically
thin scattering regime, and one can see its contribution in Fig. 1
(dotted line).

\section{Pair production}

A further effect which may strongly affect both the observed 
spectrum and the dynamics of the fireball is the production of
electron--positron pairs through photon-photon interactions.
Let us thus consider in turn the role of scattered and funnel
radiation as seed photons for this process.

\subsection{Interaction among photons in the beam}

The threshold energy for interaction between photons of energies $x$
and $x_T$ is $x_T > 2/[x(1-\cos\theta)]\sim 4\Gamma^2/x$, 
 where all quantities are calculated in the observer frame.
The
latter expression takes into account that the high energy photons
produced by the Compton drag are highly collimated, within a typical
angle $\sin\theta\sim 1/\Gamma$.
As the bulk of the scattered photons have energies $x\sim 2\Gamma^2
(3\Theta)$, pair production would occur if $\Gamma\Theta > 1/3$.

However this also implies that the scattering process
is in the Klein Nishina
regime, and we can therefore conclude that photon--photon collisions
among photons in the beam can only affect the high energy tail of the
spectrum produced at each radius, while the emission at the peak is
unaltered. We therefore neglect this effect.

\subsection{Interaction between beam photons and funnel radiation}

The interaction between the $\gamma$--rays produced by the Compton
drag process and photons emitted by the funnel walls would occur at 
large angles, resulting in an average energy threshold $x_T>1/x$.
Since $x\le\Gamma_0$, this absorption mechanism would be significant
as long as the funnel walls produce a sufficient number of photons 
with energies $x_T > 1/\Gamma_0$.

Let us then estimate the photon--photon optical depth
$\tau_{\gamma\gamma}$, by integrating the product of the
photon--photon cross section $\sigma_{\gamma\gamma}(x,x_T)$ and the
photon density above threshold $n_\gamma(x)$ over the $\gamma$--ray
path, i.e. from the site of creation, $z_1$, to infinity, and over the
photon energies:
\begin{equation}
\tau_{\gamma\gamma}(z_1,x) \, =\, 
\int_{x_T}^{\infty}dx^\prime\int_{z_1}^{\infty}
\sigma_{\gamma\gamma}(x^\prime,x) n_\gamma(z, x^\prime) dz.
\end{equation}
Since $\sigma_{\gamma\gamma}(x^\prime,x)$ is peaked at the threshold
energy, equation (11) can be simplified (Svensson 1984, 1987) as
\begin{eqnarray}
\tau_{\gamma\gamma}(z_1, x) \, = {\sigma_T \over 5 m_ec^2} 
\int_{z_1}^{\infty} x_T U(z, x_T) dz,
\end{eqnarray}
where $U(z, x_T) = m_e c^2 n_\gamma(z, x_T)$ is the photon energy 
density at threshold, at the location $z$, i.e. 
\begin{equation}
U(z_1,x_T)\, =\, {8\pi h\over c^3}\, \left( {m_ec^2\over h}\right)^4
\, { x_T^3 \over \exp[x_T/\Theta(z_1)] -1}.
\end{equation}
The radiation flux produced at the location $z_1$ is then decreased
by the factor $\exp[-\tau_{\gamma\gamma}(z_1, x)]$ while crossing the 
funnel.

The absorbed radiation will be reprocessed by the pairs,
and re--distributed in energy.
Each electron and positron will have an energy $\gamma\sim x/2$
at birth, and will cool due to the Compton drag process.
The positrons will then annihilate in collisions with the electrons 
in the fireball, producing a Doppler blueshifted annihilation line at
$x\sim \Gamma$.
We have neglected these reprocessing mechanisms, since, as can be seen 
in Fig. 1, the amount of energy absorbed in $\gamma$--$\gamma$ collisions
is small, amounting to a few per cent at most.

\begin{figure}
\vskip -0.5 true cm
\psfig{figure=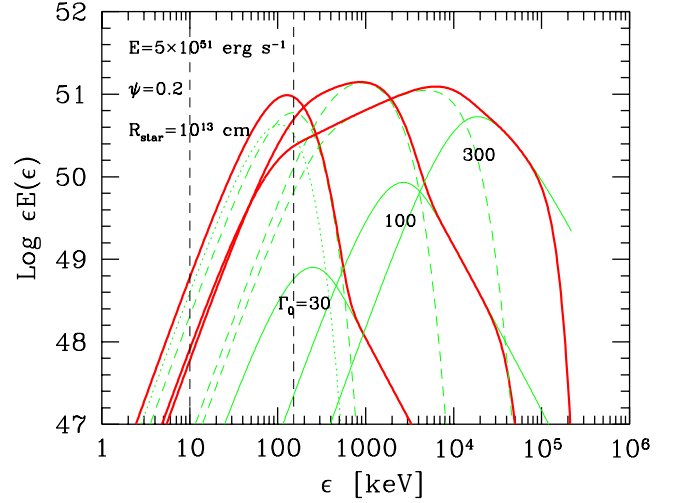,width=10cm}
\vskip -2.5 true cm
\caption{{Example of spectra produced by Compton drag.
The thick solid lines correspond to the sum of the radiation produced
inside the funnel (thin solid lines) and outside it (dashed and dotted lines).
The thin solid lines at the highest energies correspond
to the emission neglecting photon--photon absorption, to show the
importance of this process.
The dotted line (only shown for the $\Gamma_0=30$ case) is
the spectrum produced by the fireball once it becomes optically thin.
The model parameters are for all cases:
$E_f=5\times 10^{51}$ erg; $\psi=0.2$; $b=0.5$;
$g=2$; $z_*=10^{13}$ cm and $T_*=3\times 10^5$ K.
The three cases differ for the assumed initial bulk Lorentz factor
and $z_0$, i.e. $\Gamma_0=30$, $100$, $300$ and
$z_0=3\times 10^8$, $10^9$, $3\times 10^9$ cm, respectively.
The two vertical dashed lines mark 10 and 150 keV, the range of the
foreseen hard X--ray detector onboard the Swift mission.}
\label{figuno}}
\end{figure}

\begin{figure}
\psfig{figure=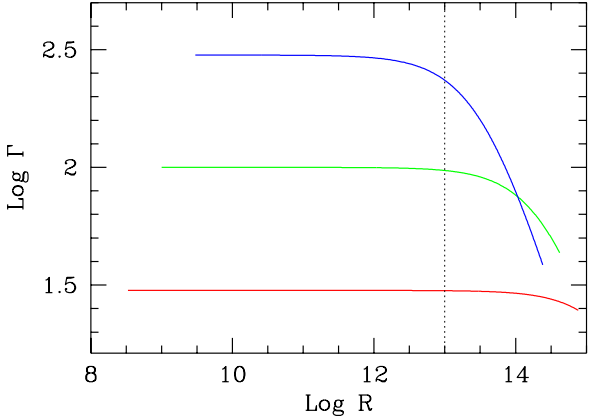,width=9cm}
\vskip -3 true cm
\caption{{The profile of the bulk Lorentz factor $\Gamma$  
corresponding to the cases shown in Fig.~1. 
The vertical dotted line marks $10^{13}$ cm, the top of the funnel.}
\label{figdue}}
\end{figure}

\begin{figure}
\vskip -0.5 true cm
\psfig{figure=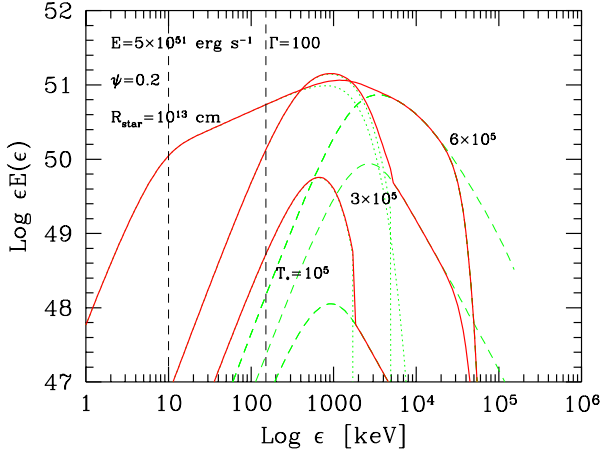,width=9cm}
\vskip -2 true cm
\caption{Spectra produced by Compton drag
for three different choices of the temperature
of the surface of the massive star (as labeled).
For all cases 
$E_f=5\times 10^{51}$ erg; $\psi=0.2$; $b=0.5$;
$g=2$; $z_*=10^{13}$ cm, $\Gamma_0=100$ and $z_0=10^9$ cm.}
\end{figure}

\section{The spectrum}

The observed total spectrum can be computed by integrating equations 
(3) and (4) over $z$, taking into account photon--photon absorption.
The contribution produced within the star is given by:  
\begin{eqnarray}
E(x) &=& \pi^2 \psi^2 m_ec^2 \left({m_e c\over h}\right)^3  
\int_{z_0}^{z_*}{z^2 \over \Gamma^6}\, 
{ x^3 e^{-\tau_{\gamma\gamma}(z, x)}\over e^{x/\Theta_c}-1}
\, dz;
\nonumber \\ &\, &  \qquad \qquad  \qquad\qquad \qquad \,\,
z_0<z<z_*, 
\end{eqnarray}
while beyond $z_*$ the number of target photons able to interact with 
high energy $\gamma$--rays to produce pairs is negligible, and thus,
ignoring photon--photon absorption, we obtain:
\begin{eqnarray}
E(x)&=&\pi^2 \psi^2 m_ec^2 \left({m_e c\over h}\right)^3  
\int_{z_*}^{z_T}{z^2 (z/z_*)^{-g} x^3\over \Gamma^6(e^{x/\Theta_{c,*}}-1) }
dz; 
\nonumber \\ 
&&  \qquad \qquad  \qquad\qquad \quad \,\,
 z_*<z<z_T.
\end{eqnarray}
In Fig.~1 we show three examples of the predicted spectrum
corresponding to different values of the initial bulk Lorentz
factors.
To illustrate the main features of the model and
the importance of photon--photon absorption, this is 
calculated both with and without the photon--photon absorption
term. 
Together with the total spectrum, the separate  
contributions for $z<z_*$ and for $z_T<z<z_*$ are reported.
In Fig. 2 we show the corresponding $\Gamma$ profiles.
The effect of the star surface temperature (and of the entire funnel, since the
parameter $b$ is assumed to be the same for all cases) can be clearly seen 
in Fig. 3.
Note the $\nu^{-1/2}$ power law shape in the X--ray band for the
high temperature case.
The extension of this power law branch depends on the value of $g$. 
In the case shown ($g=2$) the radiation energy density outside the funnel
remains sufficiently large to cause the deceleration of the fireball,
and this is responsible for the power law tail between 10 and 100 keV.
For larger $g$ the extension of this power law would decrease.
This effect can also be seen for the high $\Gamma_0$ case in Fig. 1.

In order to determine the general features of the predicted spectrum
and thus assess its robustness against the parameters of the model, we
also derived analytical (although approximated) expressions for the
spectral energy distribution.

\subsection{Analytical approximations}

First, let us approximate the blackbody spectral form with its 
Rayleigh--Jeans part, and let us neglect photon--photon absorption.
In this case, for $x<6\Theta\Gamma^2$ we have:
\begin{eqnarray}
dE(z,x) \, &\propto& \, {T\over \Gamma^4} z^2 dz;
\quad \quad {\rm for}\,\, z_o<z<z_*  \nonumber \\
&\propto &\, {T\over \Gamma^4} z^{2-g} dz;
\quad {\rm for}\,\, z_*<z<z_T.
\end{eqnarray}

Three regimes occur at different distances:

\noindent
${\mathbf z_0<z<z_d}$: ---
in this case $\Gamma=$ const, and integration over $z$ yields:
\begin{equation}
E(x) \, \propto \, x^{-(3-3b)/b};
\quad {\rm for}\, z>z_d
\end{equation}
which, for $b=0.5$, gives $E(x)\propto x^{-3}$.

\noindent
${\mathbf z_d<z<z_*}$: ---
here $\Gamma$ decreases as $(z/z_0)^{-(3-4b)}$ and thus:
\begin{equation}
E(x) \, \propto \, x^{-3(1-b)/(6-7b)}
\quad {\rm for}\,  z_d<z<z_*;
\end{equation}
which, for $b=0.5$, results in $E(x)\propto x^{-3/5}$.

\noindent
${\mathbf z_*<z<z_T}$: ---
at these distances 
the ambient radiation energy density decreases as $(z/z_0)^{-g}$.
If $\Gamma$ remains constant (= $\Gamma_*$), the spectrum
$E(x)\propto x^2$, while, for $\Gamma$ decreasing as
$\Gamma\propto (z/z_0)^{-(3-g)}$
\begin{equation}
E(x) \, \propto \, x^{-1/2}
\quad {\rm for}\,  z_*<z<z_T, 
\end{equation}
which is independent of $g$.

In conclusion, in the case of efficient Compton drag, and
independently of the particular choice of parameters, the predicted
spectrum is always characterized (in order of decreasing energy) by: a 
steep high energy tail; a first
break flagging the deceleration of the fireball; a second break
corresponding to radiation produced at the top of the funnel -- above
which the temperature of the ambient photons remains constant; a third
break, below which the spectrum $\propto x^{-1/2}$, corresponding to
the deceleration of the fireball due to the isothermal photon bath;
and finally a fourth break, below which the spectrum $F(x)\propto x^2$.
One obtains such a hard spectrum, instead of the familiar 
slope $F(x) \propto x$ corresponding to scatterings of isotropically 
distributed electrons and seed photons, because only
the photons scattered along the forward direction are observed
\footnote{This can be seen by integrating Eq. 7.23 of Rybicki 
\& Lightman (1979), 
in the angle range [$0<\theta_1< 1/\Gamma$]}.

\section{Discussion}

If the fireball propagates in a dense photon environment the
Compton drag effect must necessarily be taken into account,
and it may even be the dominant emission mechanism,
able to decelerate the fireball without the need of internal
shocks and without invoking the build--up of large magnetic fields.

In this letter  we have shown that the predicted spectrum, rather than being
simply a black body spectrum boosted in energy, has a complex
shape, with power law segments corresponding to the decrease in 
temperature of the funnel, deceleration of the fireball, and dilution
of the radiation energy density as the fireball propagates outside
the funnel while remaining optically thick.

The general features of the predicted spectrum qualitatively 
agree with observations,
since they can explain the steep power law high energy tail,
the peak of the emission, and a hard tail in the X--ray band.
The latter feature is particularly interesting, since other models
made different predictions.
In the standard internal shock synchrotron model, in fact, 
the spectrum cannot be harder than $\nu^{1/3}$ in the thin part,
and it is very unlikely that self--absorption can take place
in the X--ray band (Granot, Piran \& Sari 2000).
This would in fact imply a huge density of relativistic particles,
making the inverse Compton effect largely dominate the total
radiation output.
This radiation would be emitted at higher and yet unobserved frequencies,
and would then worsen the already severe efficiency problem.

In the quasi--thermal Comptonization model, on the other hand, the 
typical predicted spectral shape in the X--ray band is $\propto \nu^0$,
down to the typical frequencies of the seed soft photons,
i.e. the IR--optical band (Ghisellini \& Celotti 1999; 
Meszaros \& Rees 2000).

The existing observations of a significant fraction
of burst spectra harder than $\nu^{1/3}$
(Preece et al., 1999a,b; Crider et al., 1997)
are therefore already
a challenge to existing models, and may suggest a Compton
drag origin of this portion of the spectrum.
However the situation is not already a clear--cut because,
to receive enough photons to study the spectral shape, integration
times are much longer than the dynamical time--scales of the system,
with the spectrum rapidly evolving in time.
More sensitive instruments, such as the Burst Alert Telescope (BAT,
a coded mask detector more sensitive than BATSE) onboard the
foreseen Swift satellite will probably overcome this limitation.

We must also stress that the Compton drag scenario is not alternative
to the more conventional internal shock one.  Indeed, the front of the
fireball will decelerate first, plausibly causing subsequent 
un-decelerated parts to shock even if the central engine is
working in a continuous way.  This would produce additional radiation,
either by the synchrotron and inverse Compton processes or by
quasi--thermal Comptonization, depending on the details of the
particle acceleration mechanism (see Ghisellini \& Celotti 1999).  We
then expect spectral evolution: since the latter radiation mechanisms
produce a steeper low energy tail, a hard--to--soft transition
(i.e. from $\nu^2$ to $\nu^{1/3}$ or $\nu^0$) would occur.

In this paper, we have considered the illustrative case of a single fireball
moving out through an extended stellar envelope, along a funnel which is empty
of matter but pervaded by thermal radiation from the funnel walls.  
The fireball itself (for typical parameters) remains optically 
thick until it expands beyond the stellar surface.  
A burst with complex time-structure could be modeled by a series of 
fireballs or expanding shells.  
However, in this more general case,
the later shells would suffer less drag, since not enough time may 
have elapsed
to replenish the entire funnel cavity with seed photons.
Indeed one expects the spikes to be more powerful 
the longer is the time interval 
between them, as more seed photons could pervade the cavity. 
This, besides causing internal shocks with the first shell
which has been efficiently decelerated by Compton drag,
will also result in a distribution of $\Gamma$--factors:
they will become greater on axis, where few seed photons can efficiently
Compton drag the shells, and smaller towards the border of the funnel,
where seed photons can be replenished by the funnel walls.

We plan to investigate these possibilities and their consequences on the 
associated predicted afterglows in future work.

\section*{Acknowledgments}
AC acknowledges the Italian MURST for financial support.

\end{document}